# Do agents need to sleep?

Ido Haber[1], Giulio Tononi[1,*]

[1] Wisconsin Institute for Sleep and Consciousness, Department of Psychiatry, University of Wisconsin-Madison

* *Corresponding author: gtononi@wisc.edu*

## Abstract

LLM-based agents—which increasingly operate persistently across sessions and accumulate memory through interaction—already exhibit the familiar symptoms of unchecked plasticity: semantic drift, degraded retrieval precision, propagation of stored errors, and saturation of memory capacity. Drawing on the selectionist "ringing, resonance and renormalization" (RRR) account of sleep, we propose that each of its three phases maps onto an operation feasible in digital memory architectures. Engineers building stateful agents, meanwhile, are independently rediscovering the problems sleep evolved to solve: what to forget, and when to do the forgetting. From this we derive a falsifiable prediction—that selectionist consolidation will outperform instructionist alternatives as memory stores grow large. The core claim concerns the nature of continuous learning itself: plasticity has a price, and for any system that learns continuously under finite capacity while coupled to its environment, that price is sleep.

Agents work hard, often day and night. They learn by interacting with us—storing requests, conversations, and images in memory that they constantly update. The agents we are talking about are, of course, AI systems based on large language models (LLM), such as Anthropic's Claude or OpenAI's Codex. These AI agents are becoming ubiquitous because, like us, they can interact with the environment and remember what happened across a session. They are also often thought to be indefatigable. And yet, after a while, even they start showing some strain. They may need to "compact" or "consolidate" the conversation, else they run out of memory or mix up old instructions with new ones. In fact, as some have proposed, they might need something like sleep [1,2].

Is agent sleep just a casual analogy, or does it point to something deeper? We argue that sleep is not only a biological necessity but a general requirement for any system that learns continuously under finite capacity—the price to pay for plasticity [3,4]. Biological and AI agents that learn by interacting with a changing environment face the same stability-plasticity dilemma. New encoding corrupts previously consolidated knowledge [5], while signal-to-noise ratios degrade and capacity saturates [4].

Evolution's solution is sleep, a behavioral state in which organisms partially disconnect from the environment. Disconnection allows the brain to comprehensively sample its circuits and consolidate its memories, protecting what is important and down-selecting the rest [4,6]. Despite reduced vigilance and the opportunity cost of lost foraging and mating, sleep appears universal across phyla. Sleep-like states have been demonstrated not only in mammals but in fish, insects, and worms such as C. elegans with just 302 neurons, and even in jellyfish, which lack centralized brains [7].

As LLM agents transition from stateless, single-session interactions to stateful, multi-session operation, memory management turns from convenience into necessity. The costs are already apparent. Agents accumulating memory exhibit an "experience-following" property—outputs track retrieved records in proportion to input similarity—that propagates stored errors into future tasks, so naive retention causes sustained performance decline [8]. Semantic drift—the progressive divergence of iteratively summarized memories from ground truth—accumulates with each update cycle [9]. Retrieval precision degrades as irrelevant entries proliferate, and contradictions accumulate undetected. The agent's effective intelligence declines not because its model has changed but because its memory has degraded its own signal-to-noise ratio. These are not quirks of a particular implementation but consequences of encoding experience without periodic recalibration.

Engineers building stateful AI agents are independently rediscovering sleep, as evolution did across phyla. A recent survey of agent memory systems offers a comprehensive collection of the available solutions researchers have developed [1]. Different as they are, these solutions converge on the problems sleep evolved to solve: what to forget, and when to do the forgetting. It is worth asking what biology can teach us about managing the price of plasticity—particularly since early ideas about a role of biological sleep in forgetting, or "unlearning," were themselves inspired by the realization that neural networks are subject to catastrophic interference. That realization came out of theoretical work on associative memory, where superimposing too many patterns makes a network retrieve blends of what it stored [10]—the interference problem later sharpened in connectionist networks as catastrophic forgetting [5].

The Synaptic Homeostasis Hypothesis (SHY) offers a broad framework for sleep's role in plasticity [4]. According to SHY, waking interactions with the environment drive net synaptic potentiation—the strengthening of connections that underlies much of learning. But potentiation is costly. It consumes energy, occupies space, degrades signal-to-noise ratios, and saturates learning capacity [3,4]. It also risks catastrophic interference [5]. During sleep, the brain disconnects from the environment and performs memory maintenance through an "RRR" process [6]. It comprehensively "rings" its circuits, assesses which "resonate" best, and "renormalizes" synapses, protecting coherent circuits and weakening the rest.

The evidence for sleep-dependent synaptic renormalization is extensive, from mammals to invertebrates and across methodologies [6]. The size of axon-spine interfaces increases with wakefulness and decreases with sleep in different regions of the mammalian brain [3], and synaptic pruning during sleep has been observed in fish, fruit flies, and, after learning, in the worm C. elegans [6]. The critical insight of SHY for AI agents is that these costs are inherent to plasticity, not solvable by better retrieval or larger context windows. A system that learns too readily destabilizes existing knowledge, whereas one that resists change cannot adapt. The resolution is periodic renormalization, which selectively reduces the weight of accumulated traces while preserving those likely to matter most.

The RRR process that carries out synaptic homeostasis, and with it memory consolidation, is selectionist [6]. In the cerebral cortex, groups of neurons fire spontaneously every second or so (on periods), triggering waves of activity (slow waves). After a brief silence (off periods), a different set fires, and so on throughout sleep. Ringing thus samples circuits comprehensively, without the need for specific instructions. Which circuits resonate best is read out as the intensity and duration of the ensuing activity, which depend on the number and synchrony of the recurrent loops engaged, whereas firing that falls out of synchrony dissipates. Finally, during renormalization, spike-timing-dependent plasticity weakens connections whose presynaptic spikes arrive at silent postsynaptic neurons, while synapses between co-firing neurons are protected [6]. The result is a selectionist competition, in which all synapses are subject to depression and only those in circuits that resonate survive.

In principle, the three phases could map onto operations feasible in digital memory systems. Ringing might be implemented as a stochastic traversal of stored traces—probing memory in novel combinations independent of any active query—while resonance could use cosine similarity to identify semantic alignment among co-occurring traces in a vector database. Renormalization, the most developed phase in current practice, could merge redundant entries, prune low-importance records, and fold episodic detail into higher-order abstractions [1]. What is missing is the integration of all three into a unified cycle in which stochastic traversal feeds coherence evaluation into selective compression.

A selectionist approach makes it clear why the partial disconnection of sleep is a requirement, not a quirk. During wakefulness, a system must learn from the environment with which it is interacting by up-selecting circuits that resonate well "here and now". The down-selection of memories acquired over many such interactions requires the partial disconnection on both the input and output side, so that the system can comprehensively assess its circuits with no concern for the here and now. AI agents may face similar constraints. If incoming queries modify the memory store during consolidation—adding entries, updating retrieval weights—the competitive evaluation among traces is corrupted before it completes.

Disconnection need not be all-or-nothing in either biological or digital systems. Certain marine mammals and birds can interact with the environment using one hemisphere while the other sleeps [11]. Deeper in the brain, the hippocampus can enter a sleep-like state while the cortex, and the organism, remain awake for activities such as eating or grooming that require no learning or exploration [12]. And brief "local sleep" periods induced in the cerebral cortex of awake mice suffice to carry out synaptic homeostasis [13].

AI agents might similarly adopt graded disconnection rather than full shutdown. LightMem, for instance, splits memory maintenance in two: a lightweight update applied online for responsiveness, and a deferred offline pass that reconciles near-duplicates and resolves contradictions [14]. A process analogous to local sleep could run RRR on an isolated snapshot of one memory partition while the agent keeps serving queries. Comprehensive consolidation, even so, would require pausing environmental interaction entirely.

We propose that any system of finite capacity that learns continuously while coupled to its environment requires sleep-like states: periods of partial disconnection in which it comprehensively samples its memories, preserves those that fit best, and renormalizes the rest. Of course, the claim concerns the challenge and its solution, not the implementation. Agents need not replicate the slow waves, spindles, and sharp-wave ripples of mammalian non-REM sleep, or the phasic events of REM sleep—specializations that vary across phyla.

Because agent memories are discrete and addressable, instructionist strategies are far more tractable for digital agents than for brains. Mem0, for example, has a language model compare each new fact with the most similar stored memories and issue one of four operations—ADD, UPDATE, DELETE or NOOP [15]. Each decision is local, made one fact at a time while the agent is still interacting, and imposed from outside the store rather than emerging from competition within it. Whether RRR's selectionist logic outperforms such instructionist alternatives at scale is a testable hypothesis. We predict it will. As memory stores grow large and interconnected, the combinatorial complexity of deciding what to preserve will exceed the capacity of any explicit scoring function, hand-designed or learned. If so, sleep is not an accident of biology but a necessary feature of the architecture of learning. Systems that learn from the world must periodically withdraw from it—not as a luxury, but as a condition of continued function. Plasticity has a price.